\documentclass[pra,a4paper,nofootinbib,twocolumn]{revtex4}
\usepackage{amsmath}
\usepackage{amssymb}
\usepackage{amsfonts}
\usepackage{graphicx}
\usepackage{bm}

\DeclareMathOperator{\Tr}{Tr}
\DeclareMathOperator{\Sp}{sp}
\DeclareMathOperator{\tr}{tr}
\DeclareMathOperator{\li}{li}

\begin{document}

\title{Finkel'stein  nonlinear sigma model: interplay of disorder and interaction in 2D electron systems}

\author{I.S. Burmistrov
}
\affiliation{{L.~D.~Landau Institute for Theoretical Physics, acad. Semenova av. 1-a, 142432 Chernogolovka, Russia}}

\begin{abstract}
In this paper\footnote{The paper is prepared for the special JETP issue dedicated to the 100th anniversary of I. M. Khalatnikov.} I briefly review recent theoretical results derived within the Finkel'stein  nonlinear sigma model approach for description of two-dimensional  interacting disordered electron systems. The examples include an electron system with two valleys, electrons in a double quantum well, electrons on the surface of a topological insulator, an electron system with superconducting correlations, and the integer quantum Hall effect.
\end{abstract}

\maketitle


\section{Introduction\label{Sec1}}

Constant interest to disordered electron systems is largely related with
the phenomenon of Anderson localization \cite{Anderson1958}. The most convenient way to describe this~phenomenon is the scaling theory of conductance \cite{Anderson1979} which predicts localization of all single-particle electron states in dimensions $d\leqslant 2$ and existence of Anderson
transition for $d>2$. This scaling theory has been verified by direct diagrammatic calculations of conductance in weak disorder limit
\cite{Gorkov1979,Abrahams1980}. The scaling theory allows one to study the Anderson transition by means of the field theory tools developed originally for critical phenomena: low energy effective
action and renormalization group (RG) (see Refs. \cite{Amit, ZinnJustin} for a review). For the problem of Anderson localization low energy effective action is the so-called nonlinear sigma model (NLSM) \cite{Wegner1979,Schaefer1980,Efetov1980,Juengling1980,McKane1981,Efetov1982}. It describes diffusive motion of electrons on scales larger than the mean free path as interaction of diffusive modes (so-called diffusons and cooperons). The latter leads to logarithmic divergences in $d=2$.
The review of recent progress in Anderson localization can be found e.g. in Refs. \cite{Mirlin2008,Anderson50years}.

At low temperatures electron-electron interaction plays a crucial role for phenomenon of Anderson localization. The phase coherence is destructed on long time scales due to inelastic  electron-electron scattering processes with small energy transfer (compared to temperature $T$) \cite{Thouless1977,Anderson1979b,Altshuler1982}.
In addition to the phase breaking time $\tau_\phi$, electron-electron interaction results in logarithmic temperature dependence of conductance (in $d=2$) due to virtual electron-electron scattering processes \cite{Altshuler1979c,Zala2001}. Interestingly,  contribution to the conductance due to electron-electron interaction can be of opposite sign with respect to weak localization correction. This allows to speculate on existence of metal-insulator transition in $d=2$ in the presence of interactions. Experimental indications of the transition has been observed in two-dimensional (2D) electron system in Si-MOSFET \cite{Pudalov1,prb95}. 

The first attempt to include electron-electron interaction into 
the scaling theory of Anderson transition has been performed in Ref. \cite{McMillan1981}. In spite of being purely phenomenological (and incorrect due to confusion between the thermodynamic density of states and the local density of states) the scaling theory of Ref. \cite{McMillan1981} had an important  outcome: an idea of two-parameter scaling for metal-insulator transition in the presence of interaction. The 
breakthrough for the case of interacting electrons was done by Finkel'stein in Ref. \cite{Finkelstein1} where NLSM has been derived from the underlying microscopic theory. 
With the help of RG analysis of this, so-called  Finkelstein NLSM, the scaling theory of the metal-insulator transition  for $d>2$ was established in the presence of electron-electron interaction \cite{Finkelstein2,Finkelstein3,Finkelstein4,Castellani1984,Castellani1984b,Finkelstein5}. Typically, strong electron-electron interaction (e.g. Coulomb interaction) is relevant (in the RG sense) and changes the universality class of the transition in comparison with the noninteracting case (see Refs.\cite{FinkelsteinReview,KirkpatrickBelitz} for a review).

Finkel'stein nonlinear sigma model (FNLSM) is designed to describe interaction of low-energy ($|E|, T \lesssim 1/\tau_{\rm tr}$ where $\tau_\textrm{tr}$ denotes transport elastic scattering time) diffusive modes (diffusons and cooperons) in the presence of disorder and electron-electron interaction. 
FNLSM is the field theory of the matrix field $Q$ which acts in the replica space and the space of Matsubara frequencies. 
We note that FNLSM can be also formulated on the Keldysh contour (see Ref. \cite{KamenevLevchenko} for a review).
The Hermitian matrix $Q$ satisfies the nonlinear constraint
\begin{equation}
Q^2(\bm{r})=1 .
\end{equation}
Depending on the particular problem the elements
$Q^{\alpha\beta}_{mn}(\bm{r})$ can have a matrix structure and satisfies additional constraints. The Greek indices $\alpha, \beta=1,2,\dots,N_r$ stand for the replica indices whereas the Latin indices are integers $m, n$ corresponding to Matsubara frequencies $\varepsilon_n=\pi T (2n+1)$. 

The paper is organized as follows. In Sec. \ref{Sec2},  the results for an interacting disordered 2D electron system with spin-valley interplay are reviewed. In Sec. \ref{Sec3} the results for interacting electrons on the disordered surface of topological insulator thin film are presented. In Sec. \ref{Sec4} there are the results for 
2D interacting disordered electron system with superconducting correlations. The results for a 2D interacting disordered electron system in strong magnetic field are reviewed in Sec. \ref{Sec5}. The paper is concluded with discussions and outlook (Sec. \ref{Sec6}).

\section{Spin-valley interplay in an interacting disordered 2D electron system  \label{Sec2}}

In this section the interacting disordered 2D electron system with two valleys is considered. Such situation occurs in
Si(100)-MOSFET, SiO$_2$/Si(100)/SiO$_2$ quantum well, n-AlAs quantum well, and graphene (see Ref. \cite{Burmistrov2017} for a recent review). For a sake of simplicity, we assume the presence of a weak perpendicular magnetic field $B_\perp \gtrsim \max\{1/\tau_\phi,T\}/D$ where $D$ denotes the diffusion coefficient. The perpendicular magnetic field suppresses cooperons and leaves diffusons to be the only low energy diffusive modes. Also, we assume that the temperatures are not too low such that one can neglect the intra-valley elastic scattering (see Refs. \cite{Punnoose2010a,Punnoose2010b} for discussion of the effect of the intra-valley scattering).  

\subsection{Finkel'stein nonlinear sigma model}

For low energy description of an interacting diffusons in a disordered 2D electron system with spin and valley degrees of freedom the elements of the matrix field $Q^{\alpha\beta}_{mn}(\bm{r})$ are $4\times 4$  matrices in the spin and valley subspaces.  The action of FNLSM is given by two terms:
\begin{equation}
\mathcal{S} = \mathcal{S}_\sigma +\mathcal{S}_F .
\label{Ch1:Sstart:ISB}
\end{equation}
Here the first term, 
\begin{equation}
\mathcal{S}_\sigma = -\frac{\sigma_{xx}}{32} \int d\bm{r}\tr (\nabla Q)^2 ,
\label{Ch1:SstartSigma:ISB}
\end{equation}
describes NLSM for the noninteracting electrons \cite{Wegner1979,Schaefer1980,Efetov1980,Juengling1980,McKane1981}. The bare value (in the field theory sense) of $\sigma_{xx}$ is 
 the dimensionless Drude conductivity $\sigma_{xx}^{(0)}=4\pi\nu_* D$ (in units $e^2/h$)
where $\nu_*=m_*/\pi$ denotes the thermodynamic density 
of states. (The effective mass $m_*$ includes Fermi-liquid corrections.) Here and afterwards it is assumed that $\sigma_{xx}\gg 1$. 

The electron-electron interaction yields additional contribution to the NLSM action \cite{Finkelstein1,Finkelstein2,Finkelstein3,Finkelstein4}:
\begin{gather}
\mathcal{S}_F = - \pi T \int d \bm{r} \Bigl  [ \sum_{\alpha n; a b} \frac{\Gamma_{ab}}{4} \tr
I_n^\alpha t_{ab} Q(\mathbf{r}) \tr I_{-n}^\alpha t_{ab} Q(\mathbf{r})\notag \\
-4 z  \tr \eta Q \Bigr ] .
 \label{Ch1:SstartF:ISB}
\end{gather}
Here $16$ matrices $t_{ab} = {\tau}_a \otimes {\sigma}_b$
($a,b=0,1,2,3$) are the generators of the $SU(4)$ group. The Pauli matrices ${\tau}_a$ (${\sigma}_a$), $a=0,1,2,3$, act on the valley (spin) indices. The quantities ${\Gamma}_{ab}$ stand for the electron-electron interaction amplitudes. The structure of the matrix ${\Gamma}_{ab}$ is established by the microscopic derivation of the FNLSM. It is convenient to use the following parametrization: ${\Gamma}_{ab} = z \gamma_{ab}$. Here the parameter $z$ is independent charge of the field theory \eqref{Ch1:Sstart:ISB} which has been introduced originally by Finkelstein in Ref. \cite{Finkelstein1}. This additional charge (in the field theory sense) allows the RG flow to be consistent with the particle-number conservation. Parameter $z$ describes non-trivial frequency renormalization in the course of RG flow. The bare value of $z$ is determined by the thermodynamic density of states: $z^{(0)}= \pi \nu_*/4$. The renormalized value of $z$ becomes temperature dependent and governs the $T$ dependence of the specific heat \cite{CasDiCas}.  

The bare values of parameters ${\gamma}_{ab}$ can be related with the Fermi-liquid interaction parameters ${F}_{ab}$: ${\gamma}^{(0)}_{ab} = - {F}_{ab}/(1+{F}_{ab})$. More precisely, ${F}_{ab}$ are zero-angle harmonics of the Fermi liquid parameters which generalize standard singlet ($F_\rho$) and triplet ($F_\sigma$) Fermi-liquid parameters to the case of $SU(4)$ symmetry. The ${F}_{ab}$ can be estimated 
via statically screened electron-electron interaction (see for example Ref. \cite{Burmistrov2011}).  

In the presence of Coulomb interaction the bare value of $\Gamma_{00}$ is related with the bare value of $z$:
$\Gamma^{(0)}_{00}=-z^{(0)}$. Under the RG flow the quantity 
$\Gamma_{00}+z$ is conserved. Therefore, in the case of Coulomb interaction the relation $\Gamma_{00}=-z$ holds under the RG flow.  

Matrices $\Lambda$, $\eta$ and $I_k^\gamma$ are defined as follows:
\begin{gather}
\Lambda^{\alpha\beta}_{nm} =
\mathrm{sign}\,n
\delta_{nm}\delta^{\alpha\beta} t_{00},\qquad
\eta^{\alpha\beta}_{nm} = n
\delta_{nm}\delta^{\alpha\beta}t_{00},\notag \\
(I_k^\gamma)^{\alpha\beta}_{nm} =
\delta_{n-m,k}\delta^{\alpha\gamma}\delta^{\beta\gamma}t_{00}.\label{Ch1:matrices_def:ISB}
\end{gather}

\subsection{$\mathcal{F}$ --- invariance}

Matrix $Q^{\alpha\beta}_{mn}$ has formally the infinite size in the Matsubara space which is impossible to handle in practice. Therefore, one needs to introduce large frequency cut-off. We assume that  the Matsubara frequency indices are restricted to the interval $-N_M\leqslant m,n \leqslant N_M-1$ where $N_M\gg 1$. Of course, at the end of computations one needs to take the limit $N_M\to \infty$. However, this limit should be defined correctly.

The global rotations of  $Q$ by matrix $\exp(i \hat\chi)$:
\begin{equation}
Q(\bm{r}) \to e^{i \hat\chi} Q(\bm{r}) e^{-i \hat\chi} ,\,\, \hat \chi = \sum_{\alpha,n} \chi^\alpha_n I^\alpha_n \label{Ch1:GenFRot:ISB}
\end{equation}
are important due to their relation with a spatially independent electric potential~\cite{Baranov1999a,KamenevAndreev}. The latter can be gauged away by suitable gauge transformation of electron operators. To make the FNLSM action \eqref{Ch1:Sstart:ISB} consistent with this $U(1)$ gauge symmetry one needs to define the limit $N_M\to \infty$ in a such way that the following relations hold~\cite{Baranov1999a}:
\begin{eqnarray}
\tr I^\alpha_n  t_{ab} e^{i\hat\chi} Q e^{-i\hat\chi} &=& \tr I^\alpha_n  t_{ab} e^{i\chi_0} Q e^{-i\chi_0}
+ 8 i n (\chi_{ab})^\alpha_{-n}\,,\notag\\
\tr \eta e^{i\hat\chi} Q e^{-i\hat\chi} &=& \tr \eta Q +
\sum_{\alpha n;ab } i n
(\chi_{ab})^\alpha_n\tr I^\alpha_n  t_{ab}
Q \notag \\
&-&  4 \sum_{\alpha n;ab}
n^2
(\chi_{ab})^\alpha_n(\chi_{ab})^\alpha_{-n} . \label{Ch1:Falg:ISB}
\end{eqnarray}
Here $\chi_0 = \sum_{\alpha} \chi_0^\alpha I_0^\alpha$. The relations~\eqref{Ch1:Falg:ISB} guarantee the $\mathcal{F}$-invariance of the FNLSM action (i.e. its invariance under global rotations \eqref{Ch1:GenFRot:ISB} with $\chi_{ab} = \chi \delta_{a0}\delta_{b0}$)  for the case of Coulomb interaction, $\Gamma_{00}=-z$. 

\subsection{One-loop RG equations}

The renormalization of the FNLSM action \eqref{Ch1:Sstart:ISB} can be studied perturbatively in $1/\sigma_{xx}$. The lowest order treatment results in the following one-loop RG equations \cite{Burmistrov2017}
\begin{align}
\frac{d \sigma_{xx}}{d y} & =- \frac{2}{\pi} \sum_{ab} f(\Gamma_{ab}/z) , \notag  \\
\frac{d \Gamma_{ab}}{dy} &= -\frac{1}{2\pi\sigma_{xx}}
 \sum_{cd;ef} \Biggl [ 
 \bigl [  \Sp (t_{cd} t_{ef} t_{ab}) \bigr ]^2\frac{\Gamma_{cd}}{8} + \left [  \mathcal{C}_{cd;ef}^{ab} \right ]^2 \notag \\
 \times & \Bigl (\frac{\Gamma_{ab}^2}{z} - \frac{(\Gamma_{ab}-\Gamma_{cd})(\Gamma_{ab}-\Gamma_{ef})}{\Gamma_{cd}-\Gamma_{ef}}\ln \frac{z+\Gamma_{cd}}{z+\Gamma_{ef}} \Bigr ) \Biggr] , \notag \\
\frac{d z}{d y} &= \frac{1}{\pi\sigma_{xx}} \sum_{ab} \Gamma_{ab} . 
\label{eq:RG0}
\end{align}
Here $f(x) = 1-(1+1/x)\ln(1+x)$ and $y= \ln L/l$ where $L$ denotes the infrared length scale (system size). The $SU(4)$ structure constants $C^{ab}_{cd;ef}$ are defined as $[t_{cd}, t_{ef}] =
\sum_{ab} C^{ab}_{cd;ef} t_{ab}$. We note that RG Eqs. \eqref{eq:RG0} were derived to the lowest order in $1/\sigma_{xx}$. However, within this approximation the dependence on interaction parameters $\Gamma_{ab}$ in Eqs. \eqref{eq:RG0} is computed exactly.

\subsection{One-loop RG for $SU(4)$ symmetric case}

The microscopic model of the two-valley electron system, e.g. in Si-MOSFET,  does not discriminate inter- and intra-valley electron-electron interactions. This leads to the following symmetric structure of the interaction matrix $\Gamma_{ab}$ \cite{BurmistrovChtchelkatchev2008}: 
\begin{equation}
\Gamma_{ab} = z \gamma_t, \quad \textrm{for} \quad 
(ab)\neq (00) .
\end{equation}
We remind that the Coulomb interaction corresponds to the relation $\Gamma_{00}=-z$. In this case the one-loop RG Eqs. \eqref{eq:RG0} transforms into the following well-known form~\cite{PF2001}:
\begin{equation}
\begin{split}
\frac{d\sigma_{xx}}{d y} &= -\frac{2}{\pi} \left
[1+15 f(\gamma_t) \right ] ,
\\ \frac{d\gamma_t}{dy} &=
\frac{(1+\gamma_t)^2}{\pi\sigma_{xx}},\\
\frac{d\ln z}{d y} &=
\frac{15\gamma_t-1}{\pi\sigma_{xx}}. 
\end{split}
\label{eq:RG1}
\end{equation}
The RG equations \eqref{eq:RG1} predict the non-monotonic dependence of $\sigma_{xx}$ on $L$ with the ultimate metallic behavior, i.e. the increase of the conductivity, as $L\to \infty$.
We note that the $SU(4)$ symmetric manifold described by  
Eqs. \eqref{eq:RG1} is unstable with respect to general RG flow \eqref{eq:RG0} \cite{Burmistrov2017}.

\subsection{One-loop RG in the presence of symmetry breaking}

The $SU(4)$ symmetry in spin-valley space can easily be  broken by external sources, e.g. by the presence of  a finite Zeeman splitting $\Delta_s$ or a nonzero valley splitting $\Delta_v$.
The latter can be controlled by the applied stress \cite{Gunawan2006,Gunawan2007}. These symmetry breaking energy scales correspond to the length scales 
\begin{equation}
L_{s,v} = \left (\frac{\sigma_{xx}}{16 z (1+\gamma_t) \Delta_{s,v}}\right )^{1/2} . 
\end{equation}

\subsubsection{$SU(4)$ symmetry breaking by spin splitting}

We assume that $\Delta_s \gg \Delta_v$ ($L_s\ll L_v$).  Then, for short length scales $l\ll L \ll L_s\ll L_v$ the symmetry breaking terms are irrelevant and the one-loop RG equations has the form of Eqs. \eqref{eq:RG1}. At intermediate length scales, $L_s\ll L \ll L_v$, one needs to take into account the effect of the Zeeman splitting.  The non-zero $\Delta_s$ results in a mass for the triplet diffusive modes, i.e. the modes with the projection of the total spin  equal $\pm 1$. This leads to the following form of matrix field relevant at the intermediate length scales, $L_s\ll L \ll L_v$: 
\begin{equation}
Q = \sum_{a=0}^3\sum_{b=0,3}t_{ab} Q_{ab} . 
\end{equation}
The corresponding elements of the electron-electron interaction matrix has the following form:
\begin{equation}
\Gamma_{00}=-z, \quad \Gamma_{03}=z\tilde{\gamma}_t, \quad 
\Gamma_{a0}=\Gamma_{a3}=z\gamma_t, \quad a=1,2,3 .
\end{equation}
We emphasize that $\tilde{\gamma}_t \neq {\gamma}_t$, generically. This can be explained as follows. The presence of non-zero $\Delta_s$ allows one to distinguish electron-electron interaction between electrons with equal or opposite spin projections.

The modified structure of the diffusive modes as well as the interaction matrix $\Gamma_{ab}$ results in the following one loop RG equations for the intermediate length scales  $l\ll L_s\ll L \ll L_v$ \cite{BurmistrovChtchelkatchev2008}:
\begin{equation}
\begin{split}
\frac{d \sigma_{xx}}{dy} &=- \frac{2}{\pi}\left [ 1+6
f(\gamma_{t}) + f(\tilde{\gamma}_{t}) \right ] , \\
\frac{d\gamma_t}{dy} &=
\frac{1+\gamma_t}{\pi\sigma_{xx}}(1+2\gamma_t-\tilde{\gamma}_t) ,\\
\frac{d\tilde{\gamma}_t}{dy} &=
\frac{1+\tilde{\gamma}_t}{\pi\sigma_{xx}}(1-6\gamma_t-\tilde{\gamma}_t) ,\\
\frac{d\ln z}{dy} &= -\frac{1}{\pi\sigma_{xx}} \left
(1-6\gamma_t-\tilde{\gamma}_t\right ).
\end{split}
\label{eq:RG2}
\end{equation}
Here the RG running scale is defined as $y=\ln L/L_s$. Since for 
$L<L_s$ the interaction parameter $\Gamma_{03}$ coincides 
with $\Gamma_{a3}$ ($a=1,2,3$), the RG equations \eqref{eq:RG2} should be supplemented by the initial condition 
\begin{equation}
\tilde{\gamma}_t(0)=\gamma_t(0) .
\label{eq:RG2:cond}
\end{equation}
The RG equations \eqref{eq:RG2} has the unstable fixed point at $\tilde{\gamma}_t=1$ and $\gamma_t=0$. However, this fixed point is inaccessible for the initial conditions \eqref{eq:RG2:cond}, $\tilde{\gamma}_t(0)=\gamma_t(0) > 0$. The typical flow of RG Eqs. \eqref{eq:RG2}  is towards $\tilde{\gamma}_t=-1$ and $\gamma_t=\infty$. Then these RG equations become equivalent to the ones for 2 independent valleys thus leading to metallic behavior of the conductivity.

\subsubsection{$SU(4)$ symmetry breaking by both spin and valley splittings}

At the largest length scales $L\gg L_v\gg L_s\gg l$ the valley splitting $\Delta_v$ becomes important as well. Since $\Delta_v$ results in a finite mass for the diffusive modes with non-zero projection of the total valley isospin, the matrix fields $Q_{10}$, $Q_{13}$, $Q_{20}$, and $Q_{23}$ disappear from the low-energy sector of the theory. Hence, the matrix field $Q$
acquires the following form
\begin{equation}
Q = \sum_{a,b=0,3}t_{ab} Q_{ab} .
\end{equation}
In this regime only four relevant interaction parameters are left: 
\begin{equation}
\Gamma_{00}=-z, \, \Gamma_{03} = z\tilde{\gamma}_t, \, \Gamma_{30}= z\gamma_t, \, \Gamma_{33}=z\hat{\gamma}_t .
\end{equation}
The appearance of a new interaction parameter $\hat{\gamma}_t$ can be argued as follows. In the presence of strong
spin and valley splittings one can distinguished interaction
between electrons with equal and opposite spin and
isospin projections. However, the RG flow conserves the difference 
$\Gamma_{33}-\Gamma_{30}$ \cite{BurmistrovChtchelkatchev2008}. Since at $L\sim L_v$ this difference is zero, the RG flow enforces the relation $\Gamma_{33}=\Gamma_{03}$, i.e. $\hat{\gamma}_t=\gamma_t$, for $L\gg L_v\gg L_s$.

The resulting one-loop RG equations at length scales $L\gg L_v\gg L_s$ becomes~\cite{BurmistrovChtchelkatchev2008}:
\begin{equation}
\begin{split}
\frac{d \sigma_{xx}}{d y} &=-
\frac{2}{\pi} \left [ 1+2 f(\gamma_{t}) +
f(\tilde{\gamma}_{t}) \right ], \\
\frac{d\gamma_t}{d y} &=
\frac{1+\gamma_t}{\pi\sigma_{xx}}
(1-2\gamma_t-\tilde{\gamma}_{t}),\\
\frac{d\tilde{\gamma}_{t}}{dy} &=
\frac{1+\tilde{\gamma}_{t}}{\pi\sigma_{xx}}(1-2\gamma_t-\tilde{\gamma}_{t}),\\
\frac{d\ln z}{d y} &= -\frac{1}{\pi\sigma_{xx}} \left
(1-2\gamma_t-\tilde{\gamma}_{t}\right ) , 
\end{split}
\label{eq:RG3}
\end{equation}
where $y=\ln L/L_v$. There exists the line of fixed points $2\gamma_t+\tilde{\gamma}_{t}=1$. Within RG Eqs. \eqref{eq:RG3} typical behavior of conductivity is of insulating type, i.e. $\sigma_{xx}$ decreases with increase of $L$.

\subsubsection{$SU(4)$ symmetry breaking in a double quantum well}

Another breaking of $SU(4)$ symmetry occurs in an interacting disordered 2D electron system in a double quantum well. In this case the low energy effective theory can be described by the same FNLSM action \eqref{Ch1:Sstart:ISB}. However, due to the presence of a difference between inter- and intra-well electron-electron interactions, the elements $\Gamma_{ab}$ becomes as follows
\cite{Burmistrov2011}
\begin{equation}
\begin{split}
\Gamma_{00} = - z, \quad \Gamma_{10}=z \tilde{\gamma}_s, \quad \Gamma_{0a}=\Gamma_{1a}=z \gamma_t, \\
\Gamma_{20}=\Gamma_{30}=\Gamma_{2a}=\Gamma_{3a}=z\gamma_v, \quad a=1,2,3 .
\end{split}
\end{equation} 
Here $\tilde{\gamma}_s$, $\gamma_t$, and $\gamma_t$ are three dimensionless parameters which describes the electron-electron interaction in the double quantum well. The first one, $\tilde{\gamma}_s$, corresponds to the short-ranged interaction between dipoles made of electrons in two different quantum wells. The parameters $\gamma_t$ and $\gamma_v$ encode the intra- and inter-well interactions in the triplet particle-hole channel.  

The one-loop RG Eqs. \eqref{eq:RG0} acquires the following form
\begin{align}
\frac{d \sigma_{xx}}{dy} &=- \frac{2}{\pi}\bigl [1+ f(\tilde{\gamma}_{s})+6 f(\gamma_t)+
8 f(\gamma_{v})  \bigr ] ,\notag\\
\frac{d\tilde{\gamma}_s}{dy} &=
\frac{1+\tilde{\gamma}_s}{\pi\sigma_{xx}}\Bigl [ 
1-6\gamma_t-\tilde{\gamma}_s +8\gamma_v+
2 h(\tilde{\gamma}_s,\gamma_v)
\Bigr ] ,\notag\\
\frac{d{\gamma}_t}{dy} &=
\frac{1+\gamma_t}{\pi\sigma_{xx}}\Bigl [1-\tilde{\gamma}_s +2\gamma_t +
h(\gamma_t,\gamma_v)
\Bigr ] ,
\label{eq:RG4}\\
\frac{d\gamma_v}{dy} &=
\frac{1}{\pi\sigma_{xx}} \Bigl [ (1+\tilde\gamma_s)(1-\gamma_v)+2\gamma_v p(\gamma_t,\gamma_v)\Bigr  ] ,
\notag\\
\frac{d\ln z}{dy} &= \frac{1}{\pi\sigma_{xx}} \Bigl
[\tilde{\gamma}_s+6\gamma_t+8\gamma_v-1 \Bigr ] , \notag
\end{align}
where $h(u,v) = 8 v(u-v)/(1+v)$ and $p(u,v) = 1-3 u+4 v$. 

For the case of a double quantum well the initial values of the interaction parameters satisfy the following inequalities:
$\gamma_t(0)\geqslant\gamma_v(0)\geqslant 0$ and $\gamma_t(0)\geqslant \tilde{\gamma}_s(0)$ (see Ref. \cite{Burmistrov2011}). Then, as one can check, the RG Eqs. \eqref{eq:RG4} conserve the corresponding inequalities: the relations  $\gamma_t\geqslant\gamma_v\geqslant 0$ and $\gamma_t\geqslant \tilde{\gamma}_s$ are satisfied under the RG flow. Also, the interaction amplitude $\gamma_t$ increases with increase of $L$. The RG Eqs. \eqref{eq:RG4} tend toward 
$\gamma_v = 0$, $\tilde{\gamma}_s =-1$ and 
$\gamma_t = \infty$ which corresponds to separate double quantum wells. The ultimate dependence of the conductivity on $L$ is of metallic type.

\section{Interacting electrons on the disordered surface of topological insulator thin film \label{Sec3}}

In this section we consider  the interacting electrons on the disordered surface of topological insulator thin film. 3D topological insulators have no conducting states in the bulk whereas their surface hosts the electron states at the Fermi level (see Refs. \cite{Hasan2010,Qi2011} for a review). The later is the consequence of the presence of spin-orbit coupling. The properties of the surface states are affected by disorder which we assume to be non-magnetic (preserve time-reversal symmetry) and spin independent. 
Since the system has  time-reversal symmetry and no spin-rotational symmetry (due to spin-orbit coupling), it belongs to the symplectic symmetry class (in accordance with Wigner-Dyson classification). This implies that the low energy diffusive modes are singlet diffusons and cooperons. The latter are responsible for the weak anti-localization effect in the symplectic ensemble. Here we consider the general case of top and bottom surfaces of the film with unequal carrier concentration subjected to different random potentials. We neglect the effect of film ends.

\subsection{Finkel'stein nonlinear sigma model}

The low energy description of an interacting electrons on the disordered surfaces of a 3D topological insulator thin film is given in terms of the FNLSM action which has the form of Eq. \eqref{Ch1:Sstart:ISB}. Now the first term in Eq. \eqref{Ch1:Sstart:ISB} describes NLSM for the two copies (top and bottom surfaces of the film) of noninteracting electrons
\begin{equation}
\mathcal{S}_\sigma = - \sum_{s=1,2} \frac{\sigma_{xx}^{(s)}}{16} \int d\bm{r}\tr (\nabla Q_s)^2 .
\label{Ch2:SstartSigma:ISB}
\end{equation}
Here $\sigma_{xx}^{(s)}$ denotes the bare conductivity at each surface; generically, $\sigma_{xx}^{(1)}\neq \sigma_{xx}^{(2)}$. Due to the presence of time-reversal symmetry, the elements of the matrix field  $Q^{\alpha\beta}_{nm}$ are the $2\times 2$ matrix in the particle-hole space (spanned by the Pauli matrices $\tau_j$). Due to presence of strong spin-orbit coupling, $Q^{\alpha\beta}_{nm}$ has no matrix structure in the spin space. 

The contribution to the FNLSM action due to the electron-electron interaction has the following form \cite{Koenig2013}:
\begin{gather}
\mathcal{S}_F = - \pi T \int d \bm{r} \Bigl  [ \sum_{\alpha n; ss^\prime} \frac{\Gamma_{ss^\prime}}{4} \tr
I_n^\alpha (1+\tau_y) Q_s(\bm{r}) 
\notag \\
\times
\tr I_{-n}^\alpha (1+\tau_y) Q_{s^\prime}(\bm{r})
-2 \sum_s z_s  \tr \eta Q_s  \Bigr ] .
 \label{Ch2:SstartF:ISB}
\end{gather}
Here the following symmetric matrix 
\begin{equation}
\Gamma_{ss^\prime} = \begin{pmatrix}
\Gamma_{11} & \Gamma_{12} \\
\Gamma_{12} & \Gamma_{22}
\end{pmatrix} 
\end{equation}
describes the intra- ($\Gamma_{11}$ and $\Gamma_{22}$) and inter-  ($\Gamma_{12}$) surface electron-electron interaction.
The parameters $z_{1,2}$ describe frequency renormalization at each surface. 

\subsection{$\mathcal{F}$ - invariance}

As we have already explained above, the global rotations of the $Q$ matrix are of crucial importance. There are two $Q$-matrices: one for the top surface and one for the bottom surface. Therefore, there are independent rotations for each matrix:
\begin{equation}
Q_s(\bm{r}) \to W_s Q_s(\bm{r}) W_s^T  ,
\label{Ch2:GenFRot:ISB1}
\end{equation}
where
\begin{equation}
W_s = 
e^{-i \hat\chi_s^T} \frac{1+\tau_y}{2} + e^{i \hat\chi_s} \frac{1-\tau_y}{2}
,\,\, \hat \chi_s = \sum_{\alpha,n} (\chi_s)^\alpha_n I^\alpha_n  .
\end{equation}
The $U(1)$ gauge symmetry is implemented by means of  the following transformation rules \cite{Koenig2013}:
\begin{gather}
\tr I^\alpha_n\frac{1+\tau_y}{2} W_s Q W_s^T = \tr I^\alpha_n\frac{1+\tau_y}{2}  Q 
- 2 i n (\chi_s)^\alpha_{n}\,,\notag\\
\tr \eta W_s Q W_s^T = \tr \eta Q +
2 \sum_{\alpha n} i n
(\chi_{s})^\alpha_{-n}\tr I^\alpha_n \frac{1+\tau_y}{2} 
Q \notag \\
-  2 \sum_{\alpha n}
n^2
(\chi_{s})^\alpha_n(\chi_{s})^\alpha_{-n} . \label{Ch2:Falg:ISB}
\end{gather}
The FNLSM action is invariant under the global $W_s$ rotations provided the following condition holds
\begin{equation}
z_s \delta_{ss^\prime}+\Gamma_{ss^\prime}=
\textrm{const} \begin{pmatrix}
1 & -1 \\
-1 & 1
\end{pmatrix}_{ss^\prime}
 .
 \label{eq:cond-ch233}
\end{equation}
These relations between interaction parameters are fulfilled in the case of Coulomb interaction. As we shall see below, the condition \eqref{eq:cond-ch233} is consistent with the RG flow, i.e. there are three independent RG invariants 
\begin{equation}
z_1+\Gamma_{11},  \, z_2+\Gamma_{22}, \,  \Gamma_{12} .
\label{eq:RG:3:cond}
\end{equation}

\subsection{One-loop RG equations}

Since there are three RG invariants, Eq. \eqref{eq:RG:3:cond}, the RG flow  is determined by four parameters only. We choose them to be $\sigma_{xx}^{(s)}$ and $\gamma_{ss} = \Gamma_{ss}/z_s$. Then the one-loop RG equations acquire the following form \cite{Koenig2013}
\begin{equation}
\begin{split}
 \frac{d\sigma_{xx}^{(1)}}{dy} & = - \frac{2}{\pi} F\left
(\gamma_{11},\frac{\sigma_{xx}^{(1)}}{\sigma_{xx}^{(2)}}\right ),\\
 \frac{d\sigma_{xx}^{(2)}}{dy} & = - \frac{2}{\pi} F\left
(\gamma_{22},\frac{\sigma_{xx}^{(2)}}{\sigma_{xx}^{(1)}}\right ), \\
 \frac{d\gamma_{11}}{dy} &= - \frac{\gamma_{11} \left (1+\gamma_{11}\right
)}{\pi \sigma_{xx}^{(1)}},  \\
\frac{d\gamma_{22}}{dy} &= - \frac{\gamma_{22} \left (1+\gamma_{22}\right
)}{\pi \sigma_{xx}^{(2)}} ,
\end{split}
\label{eq:RG5}
\end{equation}
where 
\begin{equation}
F\left (\gamma,x\right ) = \frac{1}{2}  - \frac{(1+\gamma)\ln \left [\left (1+x\right
)\left (1 + \gamma\right )\right ]}{x \left
[1+\gamma\left (1+ \frac{1}{x}\right )\right ]}\ .
\end{equation}

Equations \eqref{eq:RG5} demonstrate rich behavior. 
There exists  a single attractive fixed point at which
the intra-surface electron-electron vanishes, $\gamma_{11}=\gamma_{22}=0$. This fixed point corresponds to the strongly coupled surfaces (due to finite inter-surface interaction $\Gamma_{12}$) with super-metallic conductivities at the top and bottom surfaces,  $\sigma_{xx}^{(1)}=\sigma_{xx}^{(2)}=\infty$. 
The fixed point with $\gamma_{11}=\gamma_{22}=-1$ corresponds to the decoupled top and bottom surfaces, $\Gamma_{12}=0$, with 
conductivities exhibiting localization behavior, i.e. $\sigma_{xx}^{(s)}\to 0$ as $L\to \infty$. However, this fixed point is unstable towards inter-surface interaction. 

We mention that the nonlinear sigma model in two dimensions for the symplectic symmetry class allows one to add the topological term to the effective action. This topological term is of Wess-Zumino-Novikov-Witten type. Due to the presence of this 
 Wess-Zumino-Novikov-Witten term in the FNLSM 
there exist non-perturbative contributions to the RG equations  \cite{Koenig2013}. In particular, in the case of decoupled surfaces this topological contribution prevents the system from localization and results in appearance of critical state on the disordered surface of 3D topological insulator \cite{Ostrovsky2010}.

\section{2D interacting disordered electron system with superconducting correlations \label{Sec4}}

In this section we consider the 2D interacting disordered electron system in the presence of superconducting correlations. Such situation is realized in 
a variety of materials, e.g. in such superconducting films as amorphous Bi and Pb [\onlinecite{Haviland89,Parendo05}], MoC
[\onlinecite{Lee90}], MoGe [\onlinecite{Yazdani95}], Ta [\onlinecite{Qin06}], InO [\onlinecite{Exp-InO,Exp-InO-1,Exp-InO-3,Exp-InO-2}], NbN [\onlinecite{Exp-NbN-1,Exp-NbN-2,Exp-NbN-3}],
TiN [\onlinecite{Exp-TiN-1,Exp-TiN-2,Exp-TiN-3,BatVinBKT}], and FeSe [\onlinecite{Exp-FeSe-1,Exp-FeSe-2,Exp-FeSe-3}]. Also, 2D superconductivity was observed in such novel
materials as LaAlO$_3$/SrTiO$_3$
[\onlinecite{Caviglia2008,Ilani2014}], SrTiO$_3$ surface [\onlinecite{Kim2012,Iwasa2014}],
MoS$_2$ [\onlinecite{Ye2012,Ye2014,Taniguchi2012}], LaSrCuO surface
[\onlinecite{Bollinger2011}], and Li$_x$ZrNCl  [\onlinecite{Exp-LiZrNCl-1,Exp-LiZrNCl-2,Exp-LiZrNCl-3,Exp-LiZrNCl-4}]. The FNLSM allows to describe the properties of the system above the superconducting transition and to estimate the transition temperature $T_c$ in the presence of disorder. We mention that the FLNSM description of 2D disordered superconductor can be extended to the region below $T_c$ (see Ref. \cite{Koenig2015} for details).

\subsection{Finkel'stein nonlinear sigma model}

In the presence of superconducting correlations the elements $Q_{nm}^{\alpha\beta}$  are $4\times 4$ matrices in the particle hole and spin spaces spanned by the Pauli matrices $\tau_{a}$ and $\sigma_b$, respectively. The action \eqref{Ch1:Sstart:ISB} should be supplemented by the additional term $\mathcal{S}_C$:
\begin{equation}
\mathcal{S} = \mathcal{S}_\sigma+\mathcal{S}_F+\mathcal{S}_C .
\label{eq:action:C}
\end{equation}
Here the first term $\mathcal{S}_\sigma$ has exactly the same form as given by Eq. \eqref{Ch1:SstartSigma:ISB}. The second term $\mathcal{S}_F$ describes now interaction in the particle-hole channel only. It is given by Eq. \eqref{Ch1:SstartF:ISB} with the following interaction matrix  ($a=1,2,3$):
\begin{equation}
\begin{split}
\Gamma_{00}=\Gamma_{30}=\Gamma_s, \, &
\Gamma_{0a}=\Gamma_{3a}=\Gamma_t , \\
\Gamma_{10}=\Gamma_{20}=0, \,  & 
\Gamma_{1a}=\Gamma_{2a}=0 .
\end{split}
\end{equation}

The interaction in the particle-particle (Cooper) channel is described by the third term in the right hand side of Eq. \eqref{eq:action:C}:
\begin{equation}
\mathcal{S}_C = -\frac{\pi T}{4}  \Gamma_c \sum_{\alpha,n} \sum_{a=1,2}  \int d\bm{r} \Tr \bigl [ t_{a0} L_n^\alpha Q \bigr ] \Tr \bigl [ t_{a0} L_n^\alpha Q \bigr ] . \label{Oc}
\end{equation} 
Here we introduced the following matrix
\begin{equation}
(L_n^\alpha)^{\beta \gamma}_{km} = \delta_{k+m,n}\delta^{\alpha\beta}\delta^{\alpha\gamma} t_{00} .
\end{equation}
The matrix field $Q$ satisfies the additional (the so-called charge-conjugation) constraint:
\begin{gather}
Q^\dag = C^T Q^T C , \qquad C = i t_{12} .
\end{gather}

\subsection{$\mathcal{F}$ invariance}

The $\mathcal{F}$ invariance of the FNLSM action \eqref{eq:action:C} is realized by the same rotations as given by Eq. \eqref{Ch1:GenFRot:ISB} with $\hat \chi \sim t_{00}$. The transformation rules are given by Eqs. \eqref{Ch1:Falg:ISB} and by the relation 
\begin{gather}
\Tr L^\alpha_n t_{a0}e^{i \hat\chi} Q e^{-i \hat\chi}  = \Tr L^\alpha_n t_{a0} Q  - 8 i \sum_{m>|n|} 
\Bigl [ (\chi_{a0})^{\alpha}_m
\notag \\
-(\chi_{a0})^{\alpha}_{-m}\Bigr ].
\label{eq:Falg:C}
\end{gather}
Using Eqs. \eqref{Ch1:Falg:ISB}  and \eqref{eq:Falg:C}, one can check that the FNLSM action  \eqref{eq:action:C} is invariant under global rotations of the matrix $Q$ with $\hat \chi \sim t_{00}$ in the case of Coulomb interaction, $\Gamma_s=-z$. 

\subsection{One-loop RG equations}

The presence of interaction in Cooper channel complicates the derivation of the RG equations \cite{KirkpatrickBelitz}. Using the background field method one can derive the following one-loop RG equations \cite{Burmistrov2015}
\begin{subequations}
\begin{align}
\frac{d \sigma_{xx}}{dy} & = -\frac{2}{\pi} \Bigl ( \frac{n-1}{2} + f(\gamma_s)+n f(\gamma_t)- \gamma_c\Bigr ) , \label{eq:RG6a}
\\
\frac{d\gamma_s}{dy}  & = - \frac{(1+\gamma_s)}{\pi\sigma_{xx}} \Bigl ( \gamma_s+n \gamma_t+2\gamma_c+4\gamma_c^2\Bigr ), \label{eq:RG6b}\\
\frac{d\gamma_t}{dy}  & = - \frac{(1+\gamma_t)}{\pi\sigma_{xx}}  \Bigl ( \gamma_s-(n-2)\gamma_t \notag \\
& \hspace{2cm} -2\gamma_c \bigl (1+2\gamma_t-2 \gamma_c \bigr ) \Bigr ), \label{eq:RG6c}
\\
\frac{d\gamma_c}{dy} & =  - 2\gamma_c^2 - \frac{1}{\pi\sigma_{xx}} \Bigl [ (1+\gamma_c)(\gamma_s- n\gamma_t) - 2\gamma_c^2 \notag \\
 & \hspace{1cm}+4\gamma_c^3 + 2 n\gamma_c \Bigl (\gamma_t-\ln(1+\gamma_t)\Bigr )\Bigr ] ,
\label{eq:RG6d}
\\
\frac{d\ln z}{dy} & = \frac{1}{\pi\sigma_{xx}} \Bigl (\gamma_s+n\gamma_t+2\gamma_c +4 \gamma_c^2\Bigr ) .\label{eq:RG6e}
\end{align}
\end{subequations}

Here $n=3$ accounts for the number of triplet particle-hole diffusive modes. In the case of strong spin-orbit coupling, the triplet diffusive modes become massive and one can use RG Eqs. \eqref{eq:RG6a}-\eqref{eq:RG6e} with $n=0$ (in this case equation for $d\gamma_t/dy$ should be omitted). 

The RG equations \eqref{eq:RG6a}-\eqref{eq:RG6e} have a very rich RG flow diagram. For $n=3$ they demonstrate tendencies towards formation of ferromagnetic phase ($\gamma_t=\infty$), insulating phase ($\sigma_{xx}=0$), and superconducting phase ($\gamma_c=-\infty$). However, Eqs. \eqref{eq:RG6a}-\eqref{eq:RG6e} become uncontrollable at the onset of these phases. For example, by comparing the first and second terms in the right hand side of Eq. \eqref{eq:RG6d} one can obtain the criterium of applicability of the one-loop RG equations for description of superconducting instability: $|\gamma_c|/\pi\sigma_{xx}\ll 1$.  

In the case, of short-ranged weak interactions, i.e. for the bare values, $|\gamma_{s0}|, |\gamma_{t0}|, |\gamma_{c0}| \ll 1$, Eqs. \eqref{eq:RG6a}-\eqref{eq:RG6e} predict the enhancement of $T_c$ in spite of the presence of disorder \cite{Burmistrov2012}. 
In this case, the analysis of superconducting instability based on RG Eqs. \eqref{eq:RG6a}-\eqref{eq:RG6e} 
is equivalent to the approach based on analysis of the self-consistent equation for the superconducting order parameter \cite{Feigel'man}.

For $n=0$ and in the case of Coulomb interaction, $\gamma_s=-1$, the RG Eqs. \eqref{eq:RG6a}-\eqref{eq:RG6e} has the attractive fixed point at $\gamma_c=1/2$ and $\sigma_{xx} = 3/\pi$. Although, this fixed point is at the boarder of applicability of the one-loop RG equations the existence of a symplectic critical metal can be a general property of the model.

It is worthwhile to mention that to the lowest order in $\gamma_c$ Eqs. \eqref{eq:RG6a}-\eqref{eq:RG6e} coincide with the original results obtained by Finkelstein long ago \cite{Finkelstein5}. 
The one-loop RG Eqs. \eqref{eq:RG6a}-\eqref{eq:RG6e} are different from equations reported in Ref. \cite{Dell'Anna2013} for the case of preserved spin-rotational and time-reversal symmetries ($n=3$). The RG equations reported in Ref. \cite{Dell'Anna2013} are inconsistent with the conservation of particles \footnote{It is instructive for the experts to highlight the difference between Eqs. \eqref{eq:RG6a}-\eqref{eq:RG6e} and that of Ref. \cite{Dell'Anna2013}. First of all, the right hand side of  the RG equation for $\gamma_s$ in Ref. \cite{Dell'Anna2013} (see Eq. (A12) there) is not proportional to the factor $1+\gamma_s$ contrary to our Eq. \eqref{eq:RG6b}. This means that the Coulomb interaction, $\gamma_s=-1$, is not the fixed point of RG equations of  Ref. \cite{Dell'Anna2013} in contradiction with  the $\mathcal{F}$-invariance of the FNLSM action. Secondly, the RG equation for $\gamma_t$ of Ref. \cite{Dell'Anna2013} does not contain the term proportional to $t \gamma_c^2$, in contrast to our Eq. \eqref{eq:RG6b}. Finally, the RG equation for $\gamma_c$ in Ref. \cite{Dell'Anna2013} contains an additional term proportional to $t \gamma_c \ln(1 + \gamma_s)$
which is absent in our Eq. \eqref{eq:RG6d}. A similar term was reported by Belitz and Kirkpatrick \cite{KirkpatrickBelitz} (see Eq. (6.8g) there). This term was criticized by Finkel'stein in Ref. \cite{Finkelstein1994}: the origin of this term has been attributed to an improper treatment of the gauge invariance in the RG scheme. We note that such terms, divergent for the case of Coulomb interaction, $\gamma_s =-1$, cannot appear in the course of renormalization of $\mathcal{F}$-invariant operators, a particular example of which is the Cooper-channel interaction term $S_C$.
}.

\section{2D interacting disordered electron system in strong magnetic field\label{Sec5}}

In this section the case of 2D interacting disordered electron system in the presence of a strong magnetic field is considered. The strong magnetic field results in two effects. At first, the magnetic field breaks time reversal symmetry and polarises the electron spin such that the matrix field $Q$ has no matrix structure in spin and particle-hole spaces (no cooperons). Secondly, the presence of magnetic field and, as the consequence, non-zero Hall conductivity $\sigma_{xy}$ allows one to add the Pruisken's theta-term into the NLSM action.

\subsection{Finkel'stein nonlinear sigma model}

The low energy effective action for 2D disordered electron system in the presence of a strong perpendicular magnetic field has the following form \cite{Pruisken1983,Pruisken1984}:
\begin{equation}
\mathcal{S}_\sigma = -\frac{\sigma_{xx}}{8} \int d\bm{r}\tr (\nabla Q)^2 + \frac{\sigma_{xy}}{8} \int d\bm{r}\tr \epsilon_{jk} Q\nabla_j Q\nabla_k Q .
\label{eq:Stop2}
\end{equation}
Here $\epsilon_{jk}$ denotes the antisymmetric tensor with $\epsilon_{xy}=-\epsilon_{yx}=1$. We remind that the last term in the right hand side of Eq. \eqref{eq:Stop2} is proportional to the integer valued topological invariant. It can be written as a purely boundary term. In the presence of electron-electron interaction the effective action involves the Finkel'stein term: 
\begin{equation}
\mathcal{S}_F = - \pi T \int d \bm{r} \Bigl  [ \sum_{\alpha n} \Gamma_{s} \tr
I_n^\alpha Q(\mathbf{r}) \tr I_{-n}^\alpha Q(\mathbf{r})
-4 z  \tr \eta Q \Bigr ] .
 \label{eq:Stop3}
\end{equation}
As above, the case of Coulomb interaction corresponds to the relation $\Gamma_s=-z$.

\subsection{$\mathcal{F}$ - invariance}

The $\mathcal{F}$ invariance of the FNLSM action \eqref{eq:action:C} is realized by the same rotations as given by Eq. \eqref{Ch1:GenFRot:ISB} with 
$\hat \chi = \sum_{\alpha n}\chi_n^\alpha I_n^\alpha$ (Here $\chi_n^\alpha$ has no internal matrix structure). The transformation rules are similar to Eqs. \eqref{Ch1:Falg:ISB}:
\begin{eqnarray}
\tr I^\alpha_n  e^{i\hat\chi} Q e^{-i\hat\chi} &=& \tr I^\alpha_n  t_{ab} Q
+ 2 i n \chi^\alpha_{-n}\,,\notag\\
\tr \eta e^{i\hat\chi} Q e^{-i\hat\chi} &=& \tr \eta Q +
2 \sum_{\alpha n} i n
\chi^\alpha_n\tr I^\alpha_n
Q \notag \\
&-&  4 \sum_{\alpha n;ab}
n^2
\chi^\alpha_n\chi^\alpha_{-n} . 
\end{eqnarray}
These relations guarantee the $\mathcal{F}$-invariance of  FNLSM for the case of Coulomb interaction, $\Gamma_s=-z$. 

\subsection{Two-loop RG equations \label{Sec6a}}

Due to the simple matrix structure of the $Q$ matrix in the case of strong magnetic field the perturbative analysis of the  FNLSM action can be extended to the next (two-loop) order in $1/\sigma_{xx}$. At present, the following two-loop results are available \cite{Baranov1999b,Burmistrov2002,Burmistrov20152}:
\begin{align}
\frac{d\sigma_{xx}}{dy} & = - \frac{2 f(\gamma_s)}{\pi}  
- \frac{4 A(\gamma_s)}{\pi^2\sigma_{xx}}  , \notag \\
\frac{d\gamma_s}{dy} & = - \frac{\gamma_s(1+\gamma_s)}{\pi \sigma_{xx}} \Bigl [
1
+ \frac{1}{\pi\sigma_{xx}}
\Bigl ( c(\gamma_s)+2 \li_2(-\gamma_s)\Bigr) \Bigr ] ,
\notag \\
\frac{d\ln z}{dy} & = \frac{\gamma_s}{\pi \sigma_{xx}}
\Bigl [
1
+ \frac{1}{\pi\sigma_{xx}}
\Bigl ( c(\gamma_s)+2 \li_2(-\gamma_s)\Bigr) \Bigr ] 
\label{eq:RG6}
\end{align}
Here the function $c(\gamma)$ is defined as follows
\begin{equation}
c(\gamma) = 2 +\frac{2+\gamma}{\gamma}\li_2(-\gamma)+ \frac{1+\gamma}{2\gamma^2}\ln^2(1+\gamma) .
\end{equation}
The function $A(\gamma_s)$ is known only for two points $\gamma_s=0$ (non-interacting electrons) and $\gamma_s=-1$ (Coulomb interaction). At $\gamma_s=0$ it is known \cite{Hikami,Ostrovsky} that 
$A(0)=1/8$. In the case of Coulomb interaction the value of $A(\gamma_s)$ is as follows \cite{Burmistrov2002}:
\begin{gather}
A(-1) =\frac{1}{16}\Bigl [\frac{139}{6}+\frac{(\pi^2-18)^2}{12}+\frac{19}{2}\zeta (3) +16\mathcal{G} \notag \\
- \Bigl (44-\frac{\pi ^{2}}{2}+7\zeta (3)\Bigr ) \ln 2+\Bigl ( 16 + \frac{\pi ^2}{3} \Bigr )\ln ^{2}2
\notag \\-\frac{1}{3}\ln ^{4}2-8\li_4\left(\frac{1}{2}\right)\Bigr ] \approx 1.64 ,
\end{gather}
where $\mathcal{G} \approx 0.915 $ denotes the Catalan constant, $\zeta(x)$ stands for the Riemann zeta-function, and $\li_n(x) = \sum_{k=1}^\infty x^k/k^n$ denotes the polylogarithm. 

The RG equations \eqref{eq:RG6} predict that the fixed point, $\gamma_s=0$, corresponding to non-interacting electrons, is stable, The fixed point, $\gamma_s=-1$, which describes the case of Coulomb interaction, is unstable. For both cases the dependence of conductivity on $L$ is of insulating type.

\subsection{Non-pertubative RG equations\label{Sec6b}}

The existence of the theta-term in the NLSM action \eqref{eq:Stop2} allows for existence of topological excitations -- instantons. They result in the following non-perturbative contributions to the RG equations \cite{Burmistrov2007}:
\begin{align}
\left [\frac{d\sigma_{xx}}{dy}\right ]_{NP} & = 
- D(\gamma_s) \sigma_{xx}^2 e^{-2\pi \sigma_{xx}} \cos(2\pi\sigma_{xy}) 
 , \notag \\
 \left [\frac{d\sigma_{xy}}{dy}\right ]_{NP} & = 
- D(\gamma_s) \sigma_{xx}^2 e^{-2\pi \sigma_{xx}} \sin(2\pi\sigma_{xy}), \notag \\
\left [\frac{d\gamma_s}{dy}\right ]_{NP} & = - \gamma_s(1+\gamma_s) D_z(\gamma_s) \sigma_{xx}  
e^{-2\pi \sigma_{xx}} \cos(2\pi\sigma_{xy})  ,
\notag \\
\left [ \frac{d\ln z}{dy}\right ]_{NP} & = 
\gamma_s D_z(\gamma_s)  \sigma_{xx}  
e^{-2\pi \sigma_{xx}} \cos(2\pi\sigma_{xy})  .
\label{eq:RG7}
\end{align}
Here $D(\gamma) = 4\pi \tilde{D}(\gamma) \exp\bigl [1-4 \gamma_E f(\gamma)\bigr ]$ where  $\gamma_E \approx 0.577$ denotes the Euler constant and 
\begin{gather}
\ln \tilde{D}(\gamma) = 2\frac{1+\gamma}{\gamma}
\Biggl\{ \Bigl [ \psi\left (\frac{1+3\gamma}{\gamma}\right )
+\psi\left (\frac{1}{\gamma}\right )-1\Bigr ]
\ln(1+\gamma) \notag \\
- g\left (-\frac{1+\gamma}{\gamma}\right )-g\left (\frac{1}{\gamma}\right ) + \frac{2\gamma^2\ln 2}{2\gamma+1} 
\Biggr \} .
\end{gather}
The function $\psi(z)$ stands for the Euler di-gamma function and $g(z)$ is defined as follows
\begin{equation}
g(z) = 2 z^2 \sum_{J=0}^\infty \frac{\ln J}{J(J^2-z^2)} .
\end{equation}
The function $D_z(\gamma) = D(\gamma) m(\gamma)$ where 
\begin{gather}
m(\gamma) =2 \frac{1+\gamma}{\gamma} e^{-2\ln(1+\gamma)/\gamma}
\int_0^\gamma ds (1+s)^{-2+2/s} .
\end{gather}
We emphasize that although different components of the function $\tilde{D}(\gamma)$ has poles on the interval $-1<\gamma<0$, the function 
$\tilde{D}(\gamma)$ has no singularities.

The non-perturbative contribution to the RG equation for $d\sigma_{xx}/dy$ has opposite sign for a half-integer value of $\sigma_{xy}$. The competition of perturbative and non-perturbative contributions at a half-integer value of $\sigma_{xy}$
can produce a non-trivial fixed point at some value $\sigma_{xx}$, both for noninteracting electrons, $\gamma_s=0$, and for electrons with Coulomb interaction, $\gamma_s=-1$.

\section{Discussions and outlook \label{Sec6}}

In this paper recent advances in the Finkel'stein nonlinear sigma model approach to interacting disordered electron systems were reviewed. This field theoretical method allows to obtain a number of interesting physical results:
\begin{itemize}
\item[(i)] In the case of 2D electron system with two valleys 
(see Sec. \ref{Sec2}) FNLSM allows us to explain peculiarities of the temperature dependence of resistivity in the presence of nonzero spin and valley splittings observed in experiments on Si-MOSFET \cite{Pudalov1997,Vitkalov,Pudalov2003} and n-AlAs
quantum well \cite{Gunawan2006,Gunawan2007} as well as 
in Al$_x$Ga$_{1-x}$As/GaAs/Al$_x$Ga$_{1-x}$As double quantum well heterostructure \cite{Minkov}.

\item[(ii)] In the case of 2D electron system at the surface of 3D topological insulator the FNLSM approach allows us to develop 
the microscopic theory of electron transport and to predict instability of the critical metallic surface state towards the inter-surface interaction.

\item[(iii)] In the case of 2D electron system with superconducting correlations the FNLSM allows us to demonstrate possibility for existence of the superconductor-insulator transition within the so-called fermionic mechanism, as well as to predict the enhancement of superconducting transition temperature in the absence of Coulomb repulsion.

\item[(iv)] In the case of 2D electron system in the presence of strong magnetic field the FNLSM allows us to substantiate the  idea of absence of Anderson transition on the perturbative level even in the presence of electron-electron interaction. Also it allows us to extend systematically the instanton physics responsible for the integer quantum Hall effect to the case with a nonzero electron-electron interaction. 

\end{itemize}

The results reviewed in this paper can be extended in several directions:
 \begin{itemize}
 \item[(i)] Extension of the known perturbative RG equations to the two-loop approximation. The available two-loop results demonstrates  complicated mathematical structure of the FNLSM which prevents obtaining higher loop RG results. 
We note that at present FNLSM lacks large-N-type parameter which would allow one to solve the problem exactly. Unfortunately, the number of valleys cannot play a role of such a parameter as two-loop RG results of Ref. \cite{Punnoose2005} demonstrate. 
 
\item[(ii)] Here FNLSM was used for description of disordered electron systems with presence or absence of standard Wigner-Dyson (time-reversal and spin-rotational) symmetries. In other words, the considered here FNLSM is extension of NLSM for the symmetry classes A, AI, and AII to the case of interacting systems. In general, noninteracting NLSM for the other 7 symmetry classes \cite{Zirnbauer,AZ} can be extended to include the terms describing electron-electron interactions  \cite{Dell'Anna2006,Dell'Anna,Liao}.

\item[(iii)] As known, NLSM has a rich non-trivial behavior of scaling dimensions of operators (without and with spatial derivatives) \cite{Wegner-mult,Hoef,Wegner-mult-2} which translates into multifractal behavior of wave functions \cite{CP} and the local density of states \cite{Lerner}, as well as into broad conductance fluctuations \cite{AKL,KLY}. Recently, the multifractal behavior of the local density of states (see Refs. \cite{Burmistrov2017m} for a review) and a non-trivial behavior of scaling dimensions of operators without spatial derivatives \cite{Repin} have been extended to FNLSM. Recently, the exact symmetry relations between scaling dimensions of these operators have been proven within NLSM approach for noninteracting electrons \cite{GMZ}. In general,
such type of exact relations could exist for the scaling dimensions of corresponding operators in the presence of electron-electron interaction, i.e. within FNLSM.
 
\end{itemize}

To summarize, more than 35 years of development of Finkel'stein nonlinear sigma model demonstrates that this theory is  internally   
consistent, convenient analytical tool for study of interplay of localization and  interactions in disordered electron systems.

\begin{acknowledgments}
I am grateful to my coauthors M. Baranov, N. Chtchelkatchev, I. Gornyi, E. K\"onig, A. Levchenko, A. Mirlin, P. Ostrovsky, I. Protopopov, A. Pruisken, K. Tikhonov for fruitful collaboration on
the problems discussed in this review. I am indebted to 
A. Germanenko, D. Knyazev, A. Kuntsevich, D. de Lang, G. Minkov, L. Ponomarenko, V. Pudalov, and A. Sherstobitov for detailed discussions of their experimental results. I thank M. Feigel'man, A. Finkelstein, Y. Fominov, A. Ioselevich, Y. Makhlin, M. Skvortsov and the other members of Landau Institute for useful discussions and comments. 

\end{acknowledgments}

\end{document}